\documentclass[12pt,epsf]{article}
\begin{document}
\def\a{\alpha}
\def\b{\beta}
\def\c{\varepsilon}
\def\d{\delta}
\def\e{\epsilon}
\def\f{\phi}
\def\g{\gamma}
\def\h{\theta}
\def\k{\kappa}
\def\l{\lambda}
\def\m{\mu}
\def\n{\nu}
\def\p{\psi}
\def\q{\partial}
\def\r{\rho}
\def\s{\sigma}
\def\t{\tau}
\def\u{\upsilon}
\def\v{\varphi}
\def\w{\omega}
\def\x{\xi}
\def\y{\eta}
\def\z{\zeta}
\def\D{{\mit \Delta}}
\def\G{\Gamma}
\def\H{\Theta}
\def\L{\Lambda}
\def\F{\Phi}
\def\P{\Psi}
\def\S{\Sigma}

\def\o{\over}
\def\beq{\begin{eqnarray}}
\def\eeq{\end{eqnarray}}
\newcommand{\gsim}{ \mathop{}_{\textstyle \sim}^{\textstyle >} }
\newcommand{\lsim}{ \mathop{}_{\textstyle \sim}^{\textstyle <} }
\newcommand{\vev}[1]{ \left\langle {#1} \right\rangle }
\newcommand{\bra}[1]{ \langle {#1} | }
\newcommand{\ket}[1]{ | {#1} \rangle }
\newcommand{\EV}{ {\rm eV} }
\newcommand{\KEV}{ {\rm keV} }
\newcommand{\MEV}{ {\rm MeV} }
\newcommand{\GEV}{ {\rm GeV} }
\newcommand{\TEV}{ {\rm TeV} }
\def\diag{\mathop{\rm diag}\nolimits}
\def\Spin{\mathop{\rm Spin}}
\def\SO{\mathop{\rm SO}}
\def\O{\mathop{\rm O}}
\def\SU{\mathop{\rm SU}}
\def\U{\mathop{\rm U}}
\def\Sp{\mathop{\rm Sp}}
\def\SL{\mathop{\rm SL}}
\def\tr{\mathop{\rm tr}}

\def\IJMP{Int.~J.~Mod.~Phys. }
\def\MPL{Mod.~Phys.~Lett. }
\def\NP{Nucl.~Phys. }
\def\PL{Phys.~Lett. }
\def\PR{Phys.~Rev. }
\def\PRL{Phys.~Rev.~Lett. }
\def\PTP{Prog.~Theor.~Phys. }
\def\ZP{Z.~Phys. }

\newcommand{\drawsquare}[2]{\hbox{%
\rule{#2pt}{#1pt}\hskip-#2pt
\rule{#1pt}{#2pt}\hskip-#1pt
\rule[#1pt]{#1pt}{#2pt}}\rule[#1pt]{#2pt}{#2pt}\hskip-#2pt
\rule{#2pt}{#1pt}}

\def\vbr{\vphantom{\sqrt{F_e^i}}}
\newcommand{\fund}{\drawsquare{6.5}{0.4}}
\newcommand{\afund}{\overline{\fund}}
\newcommand{\symm}{\drawsquare{6.5}{0.4}\hskip-0.4pt%
        \drawsquare{6.5}{0.4}}
\newcommand{\asymm}{\raisebox{-3pt}{\drawsquare{6.5}{0.4}\hskip-6.9pt%
        \raisebox{6.5pt}{\drawsquare{6.5}{0.4}}}}
\newcommand{\asymmthree}{\raisebox{-7pt}{\drawsquare{6.5}{0.4}}\hskip-6.9pt%
\raisebox{-0.5pt}{\drawsquare{6.5}{0.4}}\hskip-6.9pt%
\raisebox{6pt}{\drawsquare{6.5}{0.4}}}
\newcommand{\asymmfour}{\raisebox{-10pt}{\drawsquare{6.5}{0.4}}\hskip-6.9pt%
\raisebox{-3.5pt}{\drawsquare{6.5}{0.4}}\hskip-6.9pt%
\raisebox{3pt}{\drawsquare{6.5}{0.4}}\hskip-6.9pt%
        \raisebox{9.5pt}{\drawsquare{6.5}{0.4}}}
\newcommand{\Ythrees}{\raisebox{-.5pt}{\drawsquare{6.5}{0.4}}\hskip-0.4pt%
          \raisebox{-.5pt}{\drawsquare{6.5}{0.4}}\hskip-0.4pt%
          \raisebox{-.5pt}{\drawsquare{6.5}{0.4}}}
\newcommand{\Yfours}{\raisebox{-.5pt}{\drawsquare{6.5}{0.4}}\hskip-0.4pt%
          \raisebox{-.5pt}{\drawsquare{6.5}{0.4}}\hskip-0.4pt%
          \raisebox{-.5pt}{\drawsquare{6.5}{0.4}}\hskip-0.4pt%
          \raisebox{-.5pt}{\drawsquare{6.5}{0.4}}}
\newcommand{\Ythreea}{\raisebox{-3.5pt}{\drawsquare{6.5}{0.4}}\hskip-6.9pt%
        \raisebox{3pt}{\drawsquare{6.5}{0.4}}\hskip-6.9pt
        \raisebox{9.5pt}{\drawsquare{6.5}{0.4}}}
\newcommand{\Yfoura}{\raisebox{-3.5pt}{\drawsquare{6.5}{0.4}}\hskip-6.9pt%
        \raisebox{3pt}{\drawsquare{6.5}{0.4}}\hskip-6.9pt
        \raisebox{9.5pt}{\drawsquare{6.5}{0.4}}\hskip-6.9pt
        \raisebox{16pt}{\drawsquare{6.5}{0.4}}}
\newcommand{\Yadjoint}{\raisebox{-3.5pt}{\drawsquare{6.5}{0.4}}\hskip-6.9pt%
        \raisebox{3pt}{\drawsquare{6.5}{0.4}}\hskip-0.4pt
        \raisebox{3pt}{\drawsquare{6.5}{0.4}}}
\newcommand{\Ysquare}{\raisebox{-3.5pt}{\drawsquare{6.5}{0.4}}\hskip-0.4pt%
        \raisebox{-3.5pt}{\drawsquare{6.5}{0.4}}\hskip-13.4pt%
        \raisebox{3pt}{\drawsquare{6.5}{0.4}}\hskip-0.4pt%
        \raisebox{3pt}{\drawsquare{6.5}{0.4}}}
\newcommand{\Yflavor}{\Yfund + \overline{\Yfund}} 
\newcommand{\Yoneoone}{\raisebox{-3.5pt}{\drawsquare{6.5}{0.4}}\hskip-6.9pt%
        \raisebox{3pt}{\drawsquare{6.5}{0.4}}\hskip-6.9pt%
        \raisebox{9.5pt}{\drawsquare{6.5}{0.4}}\hskip-0.4pt%
        \raisebox{9.5pt}{\drawsquare{6.5}{0.4}}}%

\baselineskip 0.7cm

\begin{titlepage}

\begin{flushright}
UT-05-06
\end{flushright}

\vskip 1.35cm
\begin{center}
{\large \bf
Conformally Sequestered SUSY Breaking \\ in Vector-like Gauge Theories  
}
\vskip 1.2cm
M.~Ibe${}^{1}$, Izawa~K.-I.${}^{1,2}$, Y.~Nakayama${}^{1}$, Y.~Shinbara${}^{1}$, and T.~Yanagida${}^{1,2}$
\vskip 0.4cm

${}^1${\it Department of Physics, University of Tokyo,\\
     Tokyo 113-0033, Japan}

${}^2${\it Research Center for the Early Universe, University of Tokyo,\\
     Tokyo 113-0033, Japan}

\vskip 1.5cm

\abstract{
 We provide, in a framework of vector-like gauge theories, concrete models
for conformal sequestering of dynamical supersymmetry (SUSY) breaking in the 
hidden sector. If the sequestering is sufficiently strong, anomaly mediation
of the SUSY breaking may give dominant contributions to the mass spectrum of SUSY
standard-model particles, leading to negative slepton masses squared. Thus,
we also consider a model with gravitational gauge mediation to circumvent the
tachyonic slepton problem in pure anomaly mediation models.
 }
\end{center}
\end{titlepage}

\setcounter{page}{2}

\section{Introduction}

It is widely believed that conformal field theory is dynamically realized
in a large class of non-abelian gauge theories with a certain number of 
matter multiplets (see Ref.\cite{CGT}).
Conformal gauge theory is very attractive
in the phenomenological point of view, since if it includes a SUSY-breaking 
sector, conformal sequestering \cite{LS,Dine} of
the SUSY breaking may occur, providing a solution to the flavor-changing 
neutral current (FCNC) problem in the supersymmetric standard model (SSM).%
\footnote{See Ref.\cite{NS} for some other phenomenological applications
of superconformal dynamics.}
It is tempting 
to consider vector-like gauge theories for the SUSY breaking,
since they are naturally 
incorporated into vector-like superconformal gauge theories,
which are relatively well understood.

In this paper we extend vector-like gauge theories for the SUSY
breaking \cite{IYIT} by adding massive
hyperquarks to turn the full high-energy theory above the mass threshold 
into conformal gauge theory. We find, however, that this simple extension
does not achieve the conformal sequestering due to the presence of an
unwanted global $U(1)$ symmetry. To eliminate the unwanted global symmetry 
we introduce non-abelian gauge interactions acting on the additional
massive hyperquarks. We find various examples realizing the 
sequestering.

We first discuss  $SP(3N+1)\times SP(N)^6$ gauge 
theories where all gauge coupling constants at the infrared fixed point
are small for $N>1$ and perturbative calculations are applicable.
We show by an explicit 
one-loop calculation that the theories have non-trivial fixed points and the 
sequestering of the SUSY-breaking effects indeed occurs. However,
the sequestering is too mild to be applied to the phenomenology, since all
the couplings are weak.
Therefore, we dwell on strongly coupled conformal gauge
theory such as an $SP(3)\times SP(1)^2$ theory in this paper.%
\footnote{We are 
unable to prove explicitly that such a theory has a non-trivial infrared
fixed point 
and the required sequestering is obtained, since gauge couplings are
all strong. 
We only state, in this paper, why we expect that is the case.} 

We also propose a Planck-suppressed gauge mediation
which circumvents the tachyonic mass problem for sleptons in anomaly 
mediation.%
\footnote{This construction is essentially independent of the above model
of conformal SUSY breaking and serves as a generic way
to make anomaly mediation phenomenologically viable.}
Owing to the gravitational nature of this gauge mediation,
the size of the gauge-mediated SUSY breaking is at most comparable
to the anomaly-mediation effects.
For the lowest messenger scale,
the total model provides a hybrid scheme \cite{Nomura} of 
anomaly \cite{Murayama} and gauge \cite{Rattazzi} mediations of SUSY breaking.

\section{Conformal SUSY breaking}

The IYIT model \cite{IYIT} for SUSY breaking is based on an $SP(N)$ 
gauge theory with $2(N+1)$ chiral superfields (hyperquarks), $Q^i_\alpha$,
in the fundamental ($2N$-dimensional) representation.%
\footnote{We adopt the notation where $SP(1) = SU(2)$.}
Here, $\alpha
=1,\cdots,2N$ and $i =1,\cdots,2(N+1)$. We introduce $(N+1)(2N+1)$
gauge singlet chiral superfields, $S_{ij} (=-S_{ji})$, and impose the
flavor $SU(2N+2)$ symmetry in the superpotential,
\begin{equation}
W= \lambda S_{ij}Q^iQ^j,
\label{eq:tree}
\end{equation}
where $S_{ij}$ are assumed to transform as an antisymmetric 
${(N+1)(2N+1)}$ representation of the flavor $SU(2N+2)$ and we omit the
color indices for simplicity. The reason why we impose the $SU(2N+2)$ 
symmetry becomes clear in the next section.
 
The effective low-energy superpotential is given by
\begin{equation}
 W_{\rm eff}= X({\rm Pf}V^{ij}-\Lambda^{2(N+1)}) + \lambda S_{ij}V^{ij},
 \label{eq:dynamical}
\end{equation}
in terms of gauge invariant low-energy degrees of freedom
$V^{ij} \sim Q^iQ^j$.
Here, $X$ is an additional chiral superfield and $\Lambda$ denotes a
dynamical scale of the $SP(N)$ gauge interaction. We see that the
superfields $S^{ij}$ have non-vanishing $F$ terms in the vacuum and the
SUSY is spontaneously broken. Notice here that the model possesses a
$U(1)_R$ symmetry in addition to the flavor $SU(2N+2)$.

\subsection{conformality}

Now let us introduce $2n_F$ massive hyperquarks, $Q'^k$, where
$k=1,\cdots, 2n_F$. The mass term is written as
\begin{equation}
W_{\rm mass} = \sum_i mQ'^iQ'^{i+n_F}.
\label{eq:mass}
\end{equation}
Here, $i$ runs from 1 to $n_F$.
Above this mass scale, the high-energy theory is an $SP(N)$ gauge theory with 
$N_F=2(N+1)+2n_F$ hyperquarks.
The SUSY $SP(N)$ gauge theory with $N_F$ hyperquarks
is expected to be 
scale-invariant in the infrared for $3(N+1)<N_F< 6(N+1)$~\cite{Seiberg}.

We check, in the following,
that the theory with the superpotential Eq.(\ref{eq:tree}) 
can also be scale-invariant in the infrared. The NSVZ beta function \cite{NSVZ}
relates the running 
of the canonical gauge coupling constant to the anomalous dimension factors, 
$\gamma_Q$ and $\gamma_{Q'}$, of the hyperquarks, $Q$ and $Q'$, as
\begin{eqnarray}
 \mu\frac{d}{d\mu} \alpha_g = 
 -\alpha_g^2\bigg[\frac{3(N+1)-(N+1)(1-\gamma_Q)-n_F(1-\gamma_{Q'})}
 {2\pi-(N+1)\alpha_g}\bigg],
 \label{eq:betaG1}
\end{eqnarray}
where $\alpha_g$ is defined in terms of the gauge coupling constant $g$ of $SP(N)$ as 
$\alpha_g=g^2/(4\pi)$ and $\mu$ denotes the renormalization scale. Here and hereafter in this section,
we neglect the masses of the hyperquarks $Q'^k$. The beta function of the Yukawa coupling constant 
in  Eq.(\ref{eq:tree}) is also given in terms of the anomalous
dimension factors of the hyperquarks, $\gamma_Q$, and of the singlet chiral fields, $\gamma_S$, by
\begin{eqnarray}
 \mu\frac{d}{d\mu} \alpha_\lambda = 
\alpha_\lambda (\gamma_S + 2\gamma_Q),
\label{eq:betaY1}
\end{eqnarray}
where $\alpha_\lambda$ is defined in terms of the Yukawa coupling constant $\lambda$ as 
$\alpha_\lambda=\lambda^2/(4\pi)$.

When the theory is scale-invariant with non-vanishing coupling constants, 
the beta functions in Eqs.(\ref{eq:betaG1}) and (\ref{eq:betaY1})
vanish. That is, we have,
at the infrared fixed point,
\begin{eqnarray}
3(N+1)-(N+1)(1-\gamma_Q)-n_F(1-\gamma_Q')=0,
 \label{eq:fixedG1}\\
 \gamma_S + 2\gamma_Q=0.
\label{eq:fixedY1}
\end{eqnarray}
These conditions determine the anomalous dimensions at the fixed point.
The anomalous dimensions at the fixed point are consistent with the unitarity of the 
theory for
\begin{eqnarray}
-1\leq \gamma_Q, 
\quad -1 \leq \gamma_{Q'}, 
\quad  0 \leq \gamma_S, 
 \label{eq:unitarityS}
\end{eqnarray}
which comes from the restriction for unitary representation of the superconformal algebra \cite{SC}:%
\footnote{
Combining Eqs.(\ref{eq:fixedY1}) and (\ref{eq:unitarityS}), we also obtain $\gamma_Q \leq 0$
and $\gamma_S \leq 2$. 
The asymptotic freedom of $SP(N)$, namely, $n_F<2(N+1)$,
results in $\gamma_{Q'}< - \gamma_Q/2 \leq 1/2$ from Eqs.(\ref{eq:fixedG1}) and (\ref{eq:unitarityS}). 
}
the above anomalous dimensions are consistent with the unitarity conditions 
for any gauge-singlet chiral multiplets such as $QQ$, $Q'Q'$, and $S$.

Notice that the vanishing of the NSVZ beta function is consistent with
the existence of the anomaly free 
$U(1)_R$ symmetry that enters in the superconformal algebra
with the charges of the matter 
fields given by $R_i = (2+ \gamma_i)/3$; $i=Q,Q',S$ \cite{SC}.
In this simple extension of the IYIT model, the anomalous dimensions cannot be determined 
uniquely from Eqs.(\ref{eq:fixedG1}) and (\ref{eq:fixedY1}), and hence, the charge assignment 
of the $U(1)_R$ is not determined only with this information.

Now, we show by a perturbative calculation that the fixed point is infrared stable. We first
see that the gauge and Yukawa coupling constants at the infrared fixed point are small 
if $N_F$ is just below $6(N+1)$, as in the case of the Banks-Zaks fixed point~\cite{BZ}. 
In this case we can obtain the anomalous dimensions at the one-loop level as
\begin{eqnarray}
\gamma_Q &=& \frac{2N+1}{2\pi} \alpha_\lambda - \frac{2N+1}{4\pi} \alpha_g,\label{eq:anomQ1}\\
\gamma_{Q' }&=& - \frac{2N+1}{4\pi} \alpha_g, \label{eq:anomQP1}\\
\gamma_S &=& \frac{2N}{2\pi} \alpha_\lambda. \label{eq:anomS1}
\end{eqnarray}
For $n_F = 2(N+1)-\varepsilon$, we determine the coupling constants at the fixed point 
from  Eqs.(\ref{eq:fixedG1}) and (\ref{eq:fixedY1}), as
\begin{eqnarray}
\alpha^*_g &=& \frac{4\pi \varepsilon}{7N^2 + 9N +2}
\bigg(\frac{3N+1}{2N+1}\bigg)
\bigg(1+ {\cal O}\bigg(\frac{\varepsilon}{N}\bigg)\bigg),\label{eq:solG1}\\
\alpha^*_\lambda &=& \frac{2\pi \varepsilon}{7N^2 + 9N +2} 
\bigg(1+ {\cal O}\bigg(\frac{\varepsilon}{N}\bigg)\bigg),
\label{eq:solY1}
\end{eqnarray}
and the one-loop approximation is justified a posteriori for small $\varepsilon/N$.%
\footnote{
A non-perturbative determination of the coupling constants
through $a$-maximization
\cite{Intriligator:2003jj}
is given in the Appendix A.
}

We can explicitly examine the infrared stability of the fixed point by considering the 
renormalizaiton group (RG) evolutions near the fixed point.
The RG equations of the small deviations,
\begin{equation}
 \D \alpha_{g}\equiv\alpha_{g}-\alpha_{g}^*,
 \quad
 \D \alpha_{\lambda}\equiv \alpha_{\lambda}-\alpha_{\lambda}^*,
\end{equation}
are given by
\begin{eqnarray}
\mu\frac{d}{d\mu} \D \alpha_g &=& \frac{\partial \beta_g}{\partial \alpha_g}\bigg|_* \D \alpha_g\
+ \frac{\partial \beta_g}{\partial \alpha_\lambda}\bigg|_* \D \alpha_\lambda,\\
 \mu\frac{d}{d\mu} \D \alpha_\lambda &=&
  \frac{\partial \beta_\lambda}{\partial \alpha_g}\bigg|_* \D \alpha_g\
+ \frac{\partial \beta_\lambda}{\partial \alpha_\lambda}\bigg|_* \D \alpha_\lambda,
\end{eqnarray}
where $\beta_{G}$ and $\beta_{\lambda}$
denote the beta functions of $\alpha_g$ and $\alpha_\lambda$ 
given by Eqs.(\ref{eq:betaG1}) and (\ref{eq:betaY1}),
respectively, and the values with the subscript
``$*$" are evaluated at the fixed point.
By using Eqs.(\ref{eq:anomQ1})-(\ref{eq:solY1}),
we find that all the eigenvalues of the coefficient 
matrix $\{\partial \beta_k/\partial \alpha_l\}$ 
are positive at the fixed point
in Eqs.(\ref{eq:solG1}) and (\ref{eq:solY1}),
where $k, l = g, \lambda$.
Therefore, the fixed point in Eqs.(\ref{eq:solG1}) and (\ref{eq:solY1}) is infrared 
stable at least against small deviations from the fixed point.%
\footnote{
Similar situations of the conformal fixed point with non-trivial Yukawa interactions 
are discussed in  Refs.~\cite{deGouvea:1998ft}.
}

\subsection{non-sequestering}\label{sec:nonseq}

We are at the point to show that the sequestering of the SUSY breaking does not occur 
due to an unwanted global $U(1)$ symmetry in this simple extension.
By following Luty and Sundrum~\cite{LS}, we consider the RG evolutions of the wave function 
renormalization factors near the fixed point,
\begin{eqnarray}
\frac{d}{dt} \D \ln Z_i = -\gamma_i+ \gamma_i^*,
\label{eq:sequester1}
\quad
\D \ln Z_i \equiv \ln Z_i +\gamma_i^* t,
\quad
t\equiv\ln(\mu/M_*),
\end{eqnarray} 
where $i=Q,Q',S$ 
and $\gamma_i^*$ are the anomalous  dimensions at the fixed point 
given by Eqs.(\ref{eq:anomQ1})-(\ref{eq:anomS1}).
Here,
$M_*$ denotes the scale where the theory enters the conformal regime 
below the reduced Planck scale $M_G\simeq 2.4 \times 10^{18}$~GeV.
The deviations from the fixed point can be parameterized
by $\D \alpha_{g}$ and $\D \alpha_{\lambda}$
which, in turn, can be expressed as%
\footnote{
Without loss of generality, we  adopt the convention of the holomorphic 
gauge coupling in Ref.\cite{LS}.
}
\begin{eqnarray}
\D \alpha_g &=&\frac{\alpha_g^2} {2\pi-(N+1)\alpha_g}\bigg|_* 
((N+1)\D \ln Z_Q+ n_F \D \ln Z_{Q'}),\\
\D \alpha_\lambda &=& -\alpha_\lambda^*  (2\D \ln Z_Q + \D \ln Z_S).
\label{eq:deviation1}
\end{eqnarray}
By using the above expressions,  we rewrite the RG equation Eq.(\ref{eq:sequester1}) as 
\begin{eqnarray}
\frac{d}{dt} \D \ln Z_i &=&
-\bigg(\frac{\partial \gamma_i}{\partial \alpha_g}\bigg)\bigg|_* \D \alpha_g
-\bigg(\frac{\partial \gamma_i}{\partial \alpha_\lambda}\bigg)\bigg|_* \D \alpha_\lambda\\
&=&-\bigg(\bigg(\frac{\partial \gamma_i}{\partial \alpha_g}\bigg) 
\frac{\alpha_g^2} {2\pi-(N+1)\alpha_g}\bigg)\bigg|_* 
((N+1)\D \ln Z_Q+ n_F \D \ln Z_{Q'})
\nonumber\\
&  & +\bigg(\bigg(\frac{\partial \gamma_i}{\partial \alpha_\lambda}\bigg)\alpha_\lambda\bigg)\bigg|_*
	(2\D \ln Z_Q + \D \ln Z_S),
\label{eq:sequester2}
\end{eqnarray}
and we define the coefficient matrix $L_{ij}$ by
\begin{eqnarray}
\frac{d}{dt} \D \ln Z_i =\sum_{j=Q,Q',S} L_{ij} \D \ln Z_j.
\label{eq:sequester2_2}
\end{eqnarray}

When all the eigenvalues of $L$ are positive, all $\D \ln Z_i$ go to zero as $t \rightarrow -\infty$
(the infrared limit) and hence the SUSY breaking is sequestered \cite{LS,Dine}.
Unfortunately, we  find that the coefficient matrix $L$ has a zero eigenvalue.
Thus, one linear combination of $\D \ln Z_i$ is constant in the course of 
the RG evolution and it is not suppressed at the infrared fixed point. We call it 
as $\D \ln \bar{Z}$.
Since the vanishing eigenvalue corresponds to the eigenvector 
 $(\D \ln Z_Q, \D \ln Z_{Q'}, \D \ln Z_S)=(1,-(N+1)/n_F,-2)$, 
 we find that the solution to the Eq.(\ref{eq:sequester2_2}) in the infrared limit is
\begin{eqnarray}
\D \ln Z_Q &\propto&(\D \ln \bar{Z})_0,  \label{eq:solQ1}\\ 
\D \ln Z_{Q'} &\propto& -\frac{N+1}{n_F}(\D \ln \bar{Z})_0, \label{eq:solQP1}\\
\D \ln Z_S &\propto&-2 (\D \ln \bar{Z})_0,  
\label{eq:solS1}
\end{eqnarray}
with an ${\cal O}(1)$ proportionality factor, 
where $(\D \ln \bar{Z})_0$ denotes the value at $t=0$.
In general, the initial value $(\D \ln \bar{Z})_0$ contains
visible sector superfields $q_i$ as weakly coupled spectators such as
\begin{eqnarray}
(\D \ln \bar{Z})_0 \supset \frac{\kappa_{ab}}{M_G^2} q_a^{\dagger}q_b,
\label{eq:mixing}
\end{eqnarray}
where $\kappa_{ab}$ denote  ${O}(1)$ coefficients.
Therefore, from Eq.(\ref{eq:solS1}), we find that the SUSY breaking effects to the visible sector 
are not sequestered.%
\footnote{In Eqs.(\ref{eq:solQ1})-(\ref{eq:solS1}), we assume that  other eigenvalues of $L$ are
positive. Even if it is not the case, the conclusion is not changed.}

The reason of our failure can be traced to the existence of a global
$U(1)$ symmetry
\cite{LS,Dine}
under
which the SUSY breaking superfield $S_{ij}$ transforms non-trivially. 
In general, when an anomaly-free (non-$R$) $U(1)$ symmetry exists, 
the charge assignment $\omega_i$
determines the eigenvector of the coefficient matrix $L_{ij}$ for a vanishing eigenvalue:
\begin{eqnarray}
\sum_{j} L_{ij} \omega_j = 0.
\end{eqnarray}
In the present case, a linear combination $\D \ln\bar{Z}$ (of $\D \ln Z_i$) remains 
constant in the infrared limit and the SUSY breaking effects are not sequestered if the SUSY 
breaking superfields have non-vanishing charges.
The eigenvector we have found above  corresponds to the charge
assignment $(\omega_Q, \omega_{Q'}, \omega_S)=(1,-(N+1)/n_F,-2)$ of an anomaly-free $U(1)$ symmetry.
Thus, in order to realize the sequestering, we should violate the global $U(1)$ symmetry under 
which the SUSY breaking superfields transform non-trivially,
provided we do not take $(\D \ln \bar{Z})_0=0$
by fine tuning. 
In the next section, we introduce  additional gauge symmetries, where the unwanted $U(1)$ symmetry 
is broken by anomaly due to the new gauge interactions.

\section{Conformally sequestered extensions}\label{sec:seq}

We introduce gauge interactions acting on the massive hyperquarks, $Q'^k$, where the unwanted
global $U(1)$ symmetry is broken by anomaly due to the new gauge interactions. We deal, 
in this section, with $SP(N)\times SP(N')^6$, $(N=3N'+1)$ gauge theory, where the former $SP(N)$
corresponds to the gauge group for the SUSY breaking and the latter $SP(N')^6$ gauge group 
is introduced to break the unwanted $U(1)$ symmetry. 
We list all the matter contents in Table~\ref{tab:content}.
We take such a large gauge group to see explicitly by a
perturbative calculation that the conformal sequestering occurs. Indeed 
all the couplings at the infrared fixed point are weak for $N'>1$ in the present model.

\begin{table}[bt]
\begin{center}
\begin{tabular}{c|ccccccc}
        &$SP(N)$&$SP(N')$&$SP(N')$&$SP(N')$&$SP(N')$&$SP(N')$&$SP(N')$\\
 \hline
$Q\times 2(N+1)$ & $\fund_{2N} $ & $\bf 1$ & $\bf 1$ & $\bf 1$ & $\bf 1$ & $\bf 1$ & $\bf 1$ \\
\hline
$Q'$ & $\fund_{2N} $ & $\fund_{2N'}$ & $\bf 1$ & $\bf 1$ & $\bf 1$ & $\bf 1$ & $\bf 1$ \\
$Q'$ & $\fund_{2N} $ & $\bf 1$ & $\fund_{2N'}$ & $\bf 1$ & $\bf 1$ & $\bf 1$ & $\bf 1$ \\
$Q'$ & $\fund_{2N} $ & $\bf 1$ & $\bf 1$ & $\fund_{2N'}$ & $\bf 1$ & $\bf 1$ & $\bf 1$ \\
$Q'$ & $\fund_{2N} $ & $\bf 1$ & $\bf 1$ & $\bf 1$ & $\fund_{2N'}$ & $\bf 1$ & $\bf 1$ \\
$Q'$ & $\fund_{2N} $ & $\bf 1$ & $\bf 1$ & $\bf 1$ & $\bf 1$ & $\fund_{2N'}$ & $\bf 1$ \\
$Q'$ & $\fund_{2N} $ & $\bf 1$ & $\bf 1$ & $\bf 1$ & $\bf 1$ & $\bf 1$ & $\fund_{2N'}$ \\
\hline
$S_{ij}$ & $\bf 1 $ & $\bf 1$ & $\bf 1$ & $\bf 1$ & $\bf 1$ & $\bf 1$ & $\bf 1$ \\
\end{tabular}
\end{center}
\caption{The matter contents in our perturbative example, $SP(N)\times SP(N')^6$,
 $(N=3N'+1)$.
 Here, the subscripts of the fundamental representations denote the dimensions of the representations.
 In terms of the $SP(N)$ gauge theory, the number of the fundamental representation is given by
  $N_F = 2 (N+1) + 6\times 2 N'=6N-2$, 
  while the number of the fundamental representation of each $SP(N')$ gauge theory is given by
  $N_F' = 2 N = 6 N' + 2$. 
 }
\label{tab:content}
\end{table}%

\subsection{conformality}

Now, we check that the theory with the extended gauge symmetry can be scale-invariant
in the infrared.
In this model, the beta functions of the $SP(N)$ gauge coupling constant $\alpha_g$, 
the $SP(N')$ gauge coupling constant $\alpha_{g'}$, and the Yukawa coupling constant 
$\alpha_{\lambda}$ in Eq.(\ref{eq:tree}) are given by
 \begin{eqnarray}
 \mu\frac{d}{d\mu} \alpha_g &=& 
 -\alpha_g^2\bigg[\frac{3(N+1)-(N+1)(1-\gamma_Q)- 6N'(1-\gamma_{Q'})}
 {2\pi-(N+1)\alpha_g}\bigg],\label{eq:betaG2}\\
  \mu\frac{d}{d\mu} \alpha_{g'} &=& 
 -\alpha_{g'}^2\bigg[\frac{3(N'+1)-N(1-\gamma_{Q'})}
 {2\pi-(N'+1)\alpha_{g'}}\bigg],\label{eq:betaGP2}\\
\mu\frac{d}{d\mu} \alpha_\lambda &=& 
\alpha_\lambda (\gamma_S + 2\gamma_Q),
\label{eq:betaY2}
\end{eqnarray}
where we have assumed that all the $SP(N')$ sectors are equivalent. Namely, we have imposed
an exchange symmetry between any two $SP(N')$'s in the $SP(N')^6$
so that the $SP(N')^6$ has a
common gauge coupling constant $\alpha_{g'}$. 
Then, by requiring all the beta functions to vanish,
we determine the anomalous 
dimensions uniquely as
\begin{eqnarray}
\gamma_Q &=&-\frac{2(N(N+1)-9N'(N'+1))}{N(N+1)} = -\frac{4}{9N'^2+9N'+2}, \label{eq:anomQ2}\\
\gamma_{Q'} &=& \frac{N-3(N'+1)}{N}=-\frac{2}{3N'+1},\label{eq:anomQP2}\\
\gamma_S &=& -2 \gamma_Q. \label{eq:anomS2}
\end{eqnarray}
Here, we have neglected the masses of the hyperquarks $Q'$.

We also determine the coupling constants at the infrared fixed point by a perturbative calculation.
The anomalous dimensions at the one-loop level are given by
\begin{eqnarray}
\gamma_Q &=& \frac{2N+1}{2\pi} \alpha_\lambda - \frac{2N+1}{4\pi} \alpha_g,\label{eq:anomQ3}\\
\gamma_{Q' }&=& - \frac{2N'+1}{4\pi} \alpha_{g'} - \frac{2N+1}{4\pi} \alpha_g, \label{eq:anomQP3}\\
\gamma_S &=& \frac{2N}{2\pi} \alpha_\lambda. \label{eq:anomS3}
\end{eqnarray}
Then, Eqs.(\ref{eq:anomQ2})-(\ref{eq:anomS3})
determine the coupling constants
at the infrared fixed point by
\begin{eqnarray}
\frac{N+1}{2\pi}\alpha_g^* &=& \frac{8(9N'+4)}{3(2N'+1)(3N'+1)^2}, \label{eq:alphaG2}\\
\frac{N'+1}{2\pi}\alpha_{g' }^*&=& \frac{12(N'+1)(3N'^2 -3 N' -2)}{(2N'+1)(3N'+2)(3N'+1)^2}, \label{eq:alphaGP2}\\
\frac{N}{\pi}\alpha_\lambda^* &=&\frac{8}{(3N'+2)(3N'+1)}.\label{eq:alphaL2}
\end{eqnarray}
We see that all the coupling constants are small and  the perturbative calculation is reliable for $N' > 1$.%
\footnote{
For $N'=1$, although the anomalous dimensions in Eq.(\ref{eq:anomS2}) satisfy
the unitarity bound Eq.(\ref{eq:unitarityS}), 
the gauge coupling constants of $SP(N')$ in Eq.(\ref{eq:alphaG2}) turns
out to be negative, 
which implies that the perturbative description is invalid.
}

The above result enables us to explicitly analyze
the infrared stability of the fixed point in the same way
as done in the previous section.
The RG equations of the small deviations
 $\D \alpha_k \equiv \alpha_k - \alpha_k^*, (k=g,g',\lambda)$ are given by
 \begin{eqnarray}
 \mu\frac{d}{d\mu} \D \alpha_k = \sum_{l=g,g',\lambda}
 M_{kl} \D \alpha_l,
\end{eqnarray}
where the coefficient matrix $M$ is defined by
\begin{eqnarray}
M_{kl}  = \frac{\partial \beta_k}{\partial \alpha_l}\bigg|_*.
\label{eq:matrixM}
\end{eqnarray}
Here,  ``$*$'' indicates the values evaluated at the fixed point.
If all the eigenvalues of the coefficient matrix are positive, the fixed point in 
Eqs.(\ref{eq:alphaG2})-(\ref{eq:alphaL2}) is infrared stable. 
In Table~\ref{tab:stability}, we show  numerical results on the eigenvalues of the matrix $M$
for the case of $N'>1$.
>From the table, we see that eigenvalues are all positive for $N'>1$.
Therefore, we find that the $SP(3N'+1)\times SP(N')^6$ gauge theory has the stable infrared fixed point
in Eqs.(\ref{eq:anomQ2})-(\ref{eq:anomS2}).

\begin{table}[tb]
\begin{center}
\begin{tabular}{c|c|c}
         & \{$a_g^*$, $a_{g'}^*$, $a_{\lambda}^*$\} &  eigenvalues of $M$\\
 \hline
 $N'=2$&       \{0.2,  0.07,  0.1\}          &\{0.5, 0.1, 0.002\}\\
 $N'=3$&        \{0.1,  0.1,  0.07\}        &\{0.2, 0.05, 0.002\}\\
 $N'=4$&        \{0.07, 0.1,  0.04\}       &\{0.1, 0.03, 0.001\}\\
  $N'=5$&       \{0.05, 0.09, 0.03\}       &\{0.1, 0.02, 0.0006\}
 \end{tabular}
\end{center}
\caption{Stability of the infrared fixed point.
$a_{g,g',\lambda}$ denote the coupling constant,
$a_g^*=\alpha_g^*(N+1)/(2\pi)$,
$a_{g'}^*=\alpha_{g'}^*(N'+1)/(2\pi)$,
and $a_{\lambda}^*=\alpha_\lambda^* N/\pi$. 
All the eigenvalues of the coefficient matrix $M$ are positive for $N'>1$.}
\label{tab:stability}
\end{table}%

\subsection{sequestering}

We now discuss the sequestering of the SUSY breaking.
The RG equations of the wave function renormalization factors $Z_i$
near the fixed 
point are given by
\begin{eqnarray}
\frac{d}{dt}\D \ln Z_i &=& -\sum_{k=g,g',\lambda}
 \bigg(\frac{\partial \gamma_i}{\partial \alpha_k}\bigg)\bigg|_* \D \alpha_k \\
 &=& \sum_{i=Q,Q',S} L_{ij} \D \ln Z_j.
 \label{eq:sequester3}
\end{eqnarray}
Here, the coefficient matrix $L$ in the second line is given
by using the following relations: 
\begin{eqnarray}
\D \alpha_g &=&\frac{\alpha_g^2} {2\pi-(N+1)\alpha_g}\bigg|_* 
((N+1)\D \ln Z_Q+ 6 N' \D \ln Z_{Q'}),\\
\D \alpha_{g'} &=&\frac{\alpha_{g'}^2} {2\pi-(N'+1)\alpha_{g'}}\bigg|_* 
N \D \ln Z_{Q'},\\
\D \alpha_\lambda &=& -\alpha_\lambda^*  (2\D \ln Z_Q + \D \ln Z_S).
\label{eq:deviation2}
\end{eqnarray}
The sequestering of the SUSY breaking is realized when all the eigenvalues of $L$ are positive.

Interestingly, as we show below, the coefficient matrix $L$ has  the same eigenvalues as the coefficient
matrix $M$ in Eq.(\ref{eq:matrixM}).
Therefore, the sequestering occurs automatically, if the infrared fixed point determined in 
Eqs.({\ref{eq:anomQ2}})-({\ref{eq:anomS2}}) is stable.
To prove that, we rewrite the conditions for the vanishing beta functions as
\begin{eqnarray}
 \sum_{j = Q,Q',S} A_{kj} \gamma_j = b_k,
\end{eqnarray}
where the coefficient matrix $A$ and the vector $b$ can be read off from Eqs.(\ref{eq:betaG2})-(\ref{eq:betaY2}) and $k=g,g',\lambda$.
Then,
we see the following relations:
\begin{eqnarray}
M_{kl} &=& \sum_{j=Q,Q',S} A_{kj} \Gamma_{jl},\\
L_{ij} &=& \sum_{k=g,g',\lambda}  \Gamma_{ik} A_{kj},
\label{lga}
\end{eqnarray}
where we have defined
\begin{eqnarray}
\Gamma_{ik}\equiv
 \bigg(\frac{\partial \gamma_i}{\partial \alpha_k}\bigg)\bigg|_*.
\end{eqnarray}
Since the coefficient matrix $A$ is invertible,
the coefficient matrices $M$ and $L$ are similar to each other,
so that they
have the same eigenvalues.
Therefore, the sequestering occurs automatically when the anomalous dimensions are
uniquely determined by the conditions for the vanishing beta functions (i.e. $A$ is invertible)
and the fixed point is infrared stable (i.e. all the eigenvalues of $M$ are positive).
Notice that this is no accident: the conformal sequestering originates from nothing 
but the attractor structure of the infrared fixed point.

In our $SP(3N'+1)\times SP(N')^6$ model, we have shown that the fixed point is determined 
from the conditions of vanishing beta functions and the fixed point is infrared stable for $N'>1$.
Thus, we have found that the sequestering is realized in our model.
Notice that the relation between the infrared stability and the sequestering holds
independently of the perturbative calculation.
Therefore, even if perturbative analysis is not applicable, we may argue that the
sequestering occurs, if the fixed point is  expected to be infrared stable. 

It should be noted here that the unwanted global $U(1)$ symmetry discussed in the previous section
is broken by anomalies of the $SP(N')^6$ gauge interactions and hence there is no conserved $U(1)$ current.
This is the reason why the matrix $M$ does not have a zero eigenvalue.

In addition to the above global $U(1)$ symmetry,  there are many unbroken global $U(1)$'s
acting on the gauge singlet superfields $S_{ij}$, which consist of the $U(1)$ subgroups
of the flavor $SU(2N+2)$ of the hyperquarks $Q^i$.
Thus, there are many linear combinations of the wave function renormalization factors 
which are not sequestered in the course of the RG evolutions to the infrared fixed point.
For example, a linear combination $\D \ln Z_{S_{12}}-\D \ln Z_{S_{34}}$ is not sequestered,
since this corresponds to the global ${U(1)\subset SU(2N+2)}$ symmetry.
Fortunately, we can make such  non-sequestered combinations vanishing by imposing the flavor 
$SU(2N+2)$ symmetry (or a sufficiently large discrete subgroup thereof)
at high energies so that
the conformal sequestering of the SUSY breaking is realized.
Namely, by assuming that the K{\" a}hler potential inducing soft masses for squarks and sleptons 
is restricted by the flavor $SU(2N+2)$ symmetry as 
\begin{equation}
\frac{\kappa_{ab}}{M_G^2}\sum_{ij}S^{\dagger}_{ij}S_{ij}q^{\dagger a}q^{b},
\label{eq:SUGRAmass}
\end{equation}
we can set the linear combinations of $\D \ln Z$'s which are not sequestered to be zero.
Then, as we have discussed, the remaining combinations of $\D \ln Z$'s are sequestered
and the squared masses of the sfermions from Eq.(\ref{eq:SUGRAmass}) are suppressed at
the infrared fixed point. 
This is the reason why we have imposed the flavor $SU(2N+2)$ symmetry in the SUSY-breaking
sector.

Finally, in the rest of this section,
we show that the sequestering is too mild in the present model
to solve the FCNC problem. In view of the Table~\ref{tab:stability},
the smallest eigenvalue $\beta'_*$ of the coefficient matrix $L$ 
(or equivalently $M$) is of the order of $10^{-3}$ for $N'\geq 2$.
Thus, the linear combination of $\D \ln Z_i$ that corresponds to the smallest eigenvalue
approaches to the fixed point very slowly, which, in turn, prevents $\D \ln Z_i$
from getting up to the fixed point immediately.
That is, in the infrared regime ($t \ll 0$), we find
\begin{eqnarray}
\D \ln Z_S(t) &\sim& e^{\beta_*' t} \D(\ln \bar{Z})_0,\\
(\D\ln\bar{Z})_0 &=& c_S(\D\ln Z_S)_0 +c_Q(\D\ln Z_Q)_0 +c_{Q'}(\D\ln Z_{Q'})_0,
\end{eqnarray}
where $\D\ln \bar{Z}$ corresponds to the eigenvector for the smallest eigenvalue, 
the subscript ``0"
indicates the value at $t=0$,
and $c_{i}$ denote numerical coefficients.
By explicit calculation, we find that the coefficients $c_{i}$
are typically ${\cal O}(0.1-0.01)$
in our perturbative models.
In order to solve the FCNC problem by sequestering, we should require $\D \ln Z_S \lsim 10^{-7}$ at
the SUSY-breaking scale~\cite{Gabbiani:1996hi}.%
\footnote{
Here, we assume that the flavor diagonal masses of the sfermions are of the 
order of $1$~TeV, which are suppressed compared to the gravitino mass of the order of
$100$~TeV (see discussions in section~\ref{sec:Phen}).
} 
Thus, without fine tuning among $\ln Z_i$, we should require $e^{\beta_*' t} \lsim 10^{-7}$.
However, since $\beta_*'$ is of the order of $10^{-3}$, it takes too long to achieve 
the sufficient sequestering.
Therefore, we find that the sufficient sequestering cannot be expected in our perturbative models.

In the perturbative examples, we have seen that the size of the ``sequestering speed'' $\beta_*'$
 is not larger than the anomalous dimensions at the fixed point.
Thus, in order to realize the sufficient sequestering (i.e. $\beta_*' = {\cal O}(1)$), we should
require that the anomalous dimensions at the fixed point are of the order one.%
\footnote{It is based on a naive expectation that the speed of the sequestering, 
$\beta_*'\sim (\partial \gamma/\partial \alpha)\alpha|_*$ or $(\partial \gamma/\partial \alpha)\alpha^2|_*$,
is not so far from $\gamma_*$ even in the strongly coupled case 
(see Eqs.(\ref{eq:sequester3})-(\ref{eq:deviation2})).}
This means that we must consider a strongly coupled conformal gauge theory.%
\footnote{Unfortunately, the strongly coupled case $N'=1$ also seems
inadequate since the anomalous dimensions are not sufficiently large.}
In the next section, we discuss such a strongly coupled theory and present the reason 
why we consider the sequestering might be also realized there, although the perturbative
calculation is not applicable.

\section{Strongly coupled $SP(3)\times SP(1)^2$ model}
\begin{table}[tb]
\begin{center}
\begin{tabular}{c|ccc}
        &$SP(N)$&$SP(N')$&$SP(N')$\\
 \hline
$Q\times 2(N+1)$ & $\fund_{2N} $ & $\bf 1$ & $\bf 1$\\
\hline
$Q'$ & $\fund_{2N} $ & $\fund_{2N'}$ & $\bf 1$\\
$Q'$ & $\fund_{2N} $ & $\bf 1$ & $\fund_{2N'}$\\
\hline
$S_{ij}$ & $\bf 1 $ & $\bf 1$ & $\bf 1$ \\
\end{tabular}
\end{center}
\caption{The matter contents in strongly coupled model $SP(N)\times SP(N')^2$, $(N=3, N'=1)$.
Here, the subscripts of the fundamental representations denote the dimensions of the representations.
In terms of the $SP(N)$ gauge theory, the number of the fundamental representation is given by
  $N_F = 2 (N+1) + 2\times 2 N'=12$, 
  while the number of the fundamental representation of each $SP(N')$ gauge theories is given by
  $N_F' = 2 N = 6$. 
 }
\label{tab:content2}
\end{table}%
In this section, we discuss $SP(3)\times SP(1)^2$ gauge theory
as an example, where $SP(3)$ corresponds 
to the gauge group for the SUSY breaking and the $SP(1)^2$ gauge group
acts on massive hyperquarks $Q'^k$.
We list the matter contents in Table~\ref{tab:content2}.
We assume that such a gauge theory
with the Yukawa interaction in Eq.(\ref{eq:tree}) has a non-trivial fixed point. Then, the anomalous 
dimensions for $Q, Q'$, and $S$ at the fixed point are determined as
\begin{eqnarray}
\gamma_Q = -1,\quad
\gamma_{Q'}=-1,\quad
\gamma_{S}=2,
\label{eq:fixed3}
\end{eqnarray}
which sit on a boundary of the unitarity bound Eq.(\ref{eq:unitarityS}).%
\footnote{
The reason we take this example is only for simplicity.
In the phenomenological point of view, we only require a large size of the
anomalous dimensions which satisfy the unitarity bound Eq.(\ref{eq:unitarityS}).
}

In Table~\ref{tab:strong}, we list some other examples of the
$SP(N)\times SP(N')^2$ gauge theories, 
which include the cases
where the perturbative analysis is marginally applicable.
In such examples with gauge symmetry structures
similar to $SP(3)\times SP(1)^2$,
we can explicitly check that the fixed points are infrared stable.
Thus, based on these results (i.e. consistency with the unitarity and the presence of similar but
calculable examples),
we expect that the fixed point with Eq.(\ref{eq:fixed3}) is infrared stable, although it is hard to check 
it by an explicit calculation.%
\footnote{We know no calculable example that has an infrared unstable 
(non-trivial) fixed point 
in the present class of $SP(N)\times SP(N')^n, (n = 1,2,\cdots)$ gauge theories.}

\begin{table}[tb]
\begin{center}
\begin{tabular}{c|c|c|c}
         & \{ $a_g^*$ ,$a_{g'}^*$ ,$a_\lambda^*$\} & 
         $\{\gamma_Q, \gamma_{Q'}, \gamma_S\}$ &eigenvalues of $M$\\
 \hline
 $SP(3)\times SP(1)^2$&      non-perturbative &\{-1, -1, 2\}&non-perturbative\\
 $SP(5)\times SP(2)^2$&      non-perturbative &\{-0.8, -0.8, 1.6\}&non-perturbative\\
 $SP(7)\times SP(3)^2$&      non-perturbative &\{-0.7, -0.7, 1.4\}&non-perturbative\\
 $SP(13)\times SP(7)^2$&      \{0.5, 0.4, 0.3\}    &\{-0.2, -0.8, 0.4\}&\{1.6, 0.7, 0.06\}\\
 $SP(20)\times SP(11)^2$&    \{0.4, 0.5,  0.2\}  &\{-0.1, -0.8, 0.2\}&\{1.0, 0.7, 0.04\}\\
 \end{tabular}
\end{center}
\caption{Fixed point in the $SP(N)\times SP(N')$ theory.
$a_{g,g',\lambda}^*$ denote the coupling constant,
$a_g=\alpha_g^*(N+1)/(2\pi)$,
$a_{g'}^*=\alpha_{g'}^*(N'+1)/(2\pi)$,
and $a_{\lambda}^*=\alpha_\lambda^*N/\pi$. 
For the calculable examples, all the eigenvalues of the coefficient matrix $M$ are positive, and hence
the fixed points are infrared stable.}
\label{tab:strong}
\end{table}%

Now, we discuss the sequestering of the SUSY breaking effects in the strongly coupled 
$SP(3)\times SP(1)^2$ model.
As argued in the previous section, if the anomalous dimensions
are determined uniquely from the conditions
for the vanishing beta functions, 
then the sequestering is equivalent to the infrared stability of the fixed point.
Hence, once we assume that the fixed point with Eq.(\ref{eq:fixed3}) is infrared stable, 
the sequestering is guaranteed.
By assuming that the ``sequestering speed'' $\beta_*'$ is
not so far from the values of $\gamma_i$ (see Eqs.(\ref{eq:sequester3})-(\ref{eq:deviation2})), 
we expect $\beta_*' ={\cal O}(1)$ in our strongly coupled model.
The flavor-changing soft masses are sufficiently suppressed by sequestering%
\footnote{
In what follows, we assume that the theory is in the vicinity of the
conformal fixed point at $M_G$, that is, we assume $M_*\simeq M_G$ 
(see below Eq.(\ref{eq:sequester1})).}
at the energy scale $\mu$ as high as
\begin{eqnarray}
\bigg(\frac{\mu}{M_G}\bigg) \lsim 10^{-{7 \over \beta_*'}} 
 \label{eq:SQscale}.
\end{eqnarray}

When the RG scale $\mu$ comes close to the physical
mass scale $m_{\rm phys}$ of $Q'$, the theory ceases to be scale
invariant and effectively becomes an asymptotically free $SP(3)$ gauge
theory of strong coupling with 8 hyperquarks $Q$, and finally 
SUSY is broken dynamically at $\mu \lsim m_{\rm phys}$. Here, the physical mass is given by
\begin{eqnarray}
 m_{\rm phys} = (mM_G^{-\gamma_{Q'}})^{1 \over 1-\gamma_{Q'}} = \sqrt{mM_G},
 \label{eq:mphys}
\end{eqnarray}
where the last equality results from
the unitarity boundary value $\gamma_{Q'}=-1$ in the present model.
Thus, the above condition  Eq.(\ref{eq:SQscale})
for the sufficient sequestering can be rewritten in terms 
of the mass scale of $Q'$ or the SUSY breaking scale $\Lambda$ as 
\begin{eqnarray}
  \bigg(\frac{\Lambda}{M_G}\bigg) \lsim \bigg(\frac{m_{\rm phys}}{M_G}\bigg) \lsim 10^{-{7 \over \beta_*'}}.
 \label{eq:SQscale2}
\end{eqnarray}
As discussed in the next section, we are interested in the case where the gravitino mass is 
of the order of $100$~TeV, which implies $\Lambda \sim 10^{11-12}$~GeV.
Thus, we claim that the above conditions can be satisfied for $\beta_*' = {\cal O}(1)$, 
and hence, the FCNC can be suppressed in the present strongly coupled model.

\section{Circumventing the tachyonic slepton problem}\label{sec:Phen}

If the sequestering of SUSY breaking occurs sufficiently, the SUSY-breaking masses for the 
squarks and sleptons become negligibly small at low energies. 
In this situation we must invoke some mechanism to transmit 
sizable SUSY-breaking effects to the visible sector of the SSM. 
The most natural candidate is anomaly mediation \cite{Murayama}. 
This mechanism is not only theoretically interesting, but also phenomenologically attractive. 
This is because the gravitino mass is expected at ${\cal O}(100)$~TeV, which provides us
with a solution to the gravitino problem~\cite{moroi,ibe}. 
However, the anomaly mediation mechanism suffers from the tachyonic slepton mass 
problem~\cite{Murayama}.

In this section we consider a Planck-suppressed gauge mediation
to remedy this phenomenological defect of the anomaly mediation.%
\footnote{
In the Appendix B, we also provide a renormalizable setup
for such a remedy.}
Let us introduce a messenger sector
which consists of chiral superfields 
$\psi$, $\bar{\psi}$, $\psi'$, and $\bar{\psi}'$.
Here, $\psi\ ,\bar{\psi}$ and $\psi'\ ,\bar{\psi}'$ transform
as vector-like representations 
under the gauge group of the SSM,
and we take them to
fit in complete $SU(5)$ GUT representations, $\mathbf{5}+\mathbf{5^*}$,
for simplicity. 

Our additional superpotential terms are given by
\begin{equation}
\delta W = {h \over M_G^2}S_{ij}Q^iQ^j\psi \bar{\psi}
 + m_m \psi \bar{\psi}' + m_m \psi' \bar{\psi},
 \label{eq:GM1}
\end{equation}
which let the SUSY breaking intact for a sufficiently large
mass parameter $m_m$.%
\footnote{In fact, $m_m \gsim m_{3/2}$ is required,
where $m_{3/2}$ denotes the gravitino mass.}
Here, $h$ denotes a coupling constant of order one and
the combination $S_{ij}Q^iQ^j$ stands just for a SUSY-breaking
superpotential term which has a non-vanishing $F$ component
(see Eq.(\ref{eq:dynamical})).
The SUSY-breaking effects are transmitted to the sfermions and 
Higgs bosons by the SSM gauge interactions
(see Ref.\cite{Tobe}).  

In the SUSY-breaking dynamics, we expect $|\vev{S}| \lsim \L$ 
\cite{Hotta}, which yields only Planck-suppressed $R$ breaking effects in the
gauge mediation.
In this case, gauginos do not obtain sizable SUSY-breaking masses
via the gauge mediation
and the gaugino spectrum is virtually
the same as in the purely anomaly-mediated one.%
\footnote{
Since the superpotential has the constant term which is required to obtain
the flat universe,
we may as well introduce an interaction term between $\psi\bar{\psi}$ and 
the constant term.
Then, the $R$ breaking effects in 
the gauge mediation possibly become sizable \cite{Tobe},
which may result in the gaugino spectrum
different from the one in the pure anomaly mediation.
The expression of the gauge mediated mass squared in Eq.(\ref{eq:GMmass})
may also be altered.
}
On the other hand, the scalar field $\phi$ obtains
the mass squared via the gauge mediation for $m_m < m_{\rm phys}$ as
\begin{eqnarray}
m_{\phi}^2 &\simeq& 2 \sum_{a=1,2,3} C_a^{\phi}
\bigg(\frac{\alpha_a}{4\pi}\bigg)^2
 \frac{|h F_S|^2}{m_m^2} \bigg({\L \over  M_G}\bigg)^4,\\
&\simeq& 18 \sum_{a=1,2,3} C_a^{\phi}\bigg(\frac{\alpha_a}{4\pi}\bigg)^2
\bigg({|h| m_{3/2} \over |\lambda| m_m}\bigg)^2 m_{3/2}^2,
\label{eq:GMmass}
\end{eqnarray}
where $C_a^{\phi} (a=1,2,3)$ is the quadratic Casimir
invariant for each gauge group relevant to the scalar $\phi$.%
\footnote{Here, we have neglected RG effects from the MSSM couplings.}
In the above equation, we have used
$\sqrt{F_S} \simeq \sqrt{\lambda}\Lambda$
and the gravitino mass $m_{3/2}$ given by
\begin{eqnarray}
m_{3/2}\simeq\frac{|F_S|}{\sqrt{3}M_G}
\simeq\frac{|\lambda| \Lambda^2}{\sqrt{3}M_G}.
\end{eqnarray}

As a result, we find that the gauge-mediated masses squared 
are comparable to the anomaly mediated ones for $m_m \sim m_{3/2}$.%
\footnote{
For $m_m \lsim m_{3/2}$, even if the total SUSY-breaking were kept intact,
messenger scalar particles would become tachyonic. Hence
we restrict ourselves to $m_m \gsim m_{3/2}$.
The choice $m_m \sim m_{3/2}$ realizes the lowest-scale model
of gauge mediation (see Ref.\cite{Nomura,Iza})
for $m_{3/2}$ of order $100$~TeV.
We note that $m_m \sim m_{3/2}$ is realized
by a relation $m_m \sim m$ of the Lagrangian parameters
in view of Eq.(\ref{eq:mphys}).}
In particular, the positive contributions to the slepton masses squared
in Eq.(\ref{eq:GMmass}),
\begin{eqnarray}
m_{\tilde{e}\,\, \rm G.M.}^2  \simeq  \frac{3}{5}18 \bigg(\frac{|h| m_{3/2}}{|\lambda| m_m}\bigg)^2
\bigg(\frac{\alpha_1}{4\pi}\bigg)^2m_{3/2}^2,
\label{eq:GMse}
\end{eqnarray}
can overwhelm the negative contribution of the anomaly-mediated mass squared,
\begin{eqnarray}
m_{\tilde{e}\,\,\rm A.M.}^2
\simeq - \frac{6}{5}\frac{33}{5}\bigg(\frac{\alpha_1}{4\pi}\bigg)^2m_{3/2}^2.
 \label{eq:AMse}
\end{eqnarray}
Therefore, we conclude that the tachyonic slepton problem is resolved
in the total model of anomaly and gauge mediation hybrid by tunning
scales of these two mediations with each other.

Note that the newly added superpotential term in Eq.(\ref{eq:GM1}) 
does not violate the global $SU(2N+2)$ symmetry which is relevant for the conformal sequestering.
Hence, we can apply the above mechanism to the conformally sequestered models.
However, we should note that in the case of conformally sequestered models,
the first term in Eq.(\ref{eq:GM1}) is also sequestered in the course of the RG evolution
from $M_{*}$ to $m_{\rm phys}$.%
\footnote{The sequestering can be seen through a field redefinition
${\tilde S}_{ij}=(1+\lambda^{-1} h \psi \bar{\psi}/M_G^2)S_{ij}$,
which turns the effects of the superpotential coupling $h$
into those of the K{\" a}hler couplings
appearing as perturbations to the renormalization factors.
}
Thus, in order to realize the sizable gauge mediation effects as in Eq.(\ref{eq:GMse}),
we need to compensate the sequestering effects by preparing the additional 
superpotential terms 
\begin{equation}
 \left( {M_* \over m_{\rm phys}} \right)^{\beta'_*}
 {h \over M_G^2}S_{ij}Q^iQ^j\psi \bar{\psi}
 + m_m \psi \bar{\psi}' + m_m \psi' \bar{\psi}
\end{equation}
at the scale $M_*$ which effectively realize the Eq.(\ref{eq:GM1}) after the conformal sequestering.
This implies that the higher-dimensional term
stems from integrating out an intermediate matter of mass
$({m_{\rm phys}/M_*})^{\beta'_*}M_G$ with Planck-suppressed coupling
to $S_{ij}Q^iQ^j$.%
\footnote{
For example, we may consider a concrete model 
by introducing extra singlet supermultiplets $X$ and $\bar{X}$ with a superpotential
\begin{eqnarray}
 \frac{h_1}{M_{G}} X SQQ + M_{X} X\bar{X} + h_2 \bar{X} \psi\bar\psi,
\end{eqnarray}
at the scale $M_{*}$.
Here, $h_{1,2}$  denote coupling constants, $M_{X}$ the mass parameter
of $X$. 
Then, after integrating out $X$ and $\bar{X}$,  we can effectively obtain 
the first term in Eq.(\ref{eq:GM1}) at the scale $m_{\rm phys}$ for $M_{X}\sim M_{G} (m_{\rm phys}/M_{*})^{\b_{*}'} ((h_{1}h_{2})/h)$.
}
In the above analysis, we have simply used Eq.(\ref{eq:GM1}) as a resultant 
effective superpotential at the scale $m_{\rm phys}$ for the conformally sequestered models.

Finally, we comment on the cosmologcal aspects of this class of models.
Since the relic density of the lightest messenger particle is too much to be consistent with 
the observation, we should require that they decay into the SSM particles at early
stage of the universe.
This is implemented by introducing small mixings between the messenger particles and the SSM
particles.
In addition, it should also be noted that there are Goldstone bosons in the SUSY breaking sector,
 which correspond to the spontaneous breaking of the global $SU(2N+2)$ symmetry.
However, those massless particles are decoupled from the thermal bath since they only couple
with the SSM particles via the Planck-supressed operator, and hence
they do not affect the history of the universe.%
\footnote{Here, we assume that the inflaton decays dominantly to the SSM particles.}
Therefore, we find that the present hybrid scheme
yields also a consistent scenario from the cosmological
point of view.

\section*{Acknowledgements}
M.I. and Y.N thank the Japan Society for the Promotion of Science for financial
support. 
The authors acknowledges the referee for useful comments.  

\appendix
\section{Anomalous dimensions from $a$-maximization}
Recently, Intriligator and Wecht proposed a powerful technique to
compute the conformal $R$ current in a certain class of
conformal field theories in four dimensions
and hence the anomalous dimensions thereof \cite{Intriligator:2003jj}.
In this appendix we use this so-called $a$-maximization method
to determine the anomalous dimensions of the fields
in the conformally extended IYIT model
beyond the Banks-Zaks approximation presented in section 3.

The $a$-maximization method simply states that the conformal $R$ current
appearing in the superconformal algebra maximizes
a particular t'Hooft anomaly
\begin{equation}
a = \mathrm{Tr}(3R^3 - R),
\end{equation}
which is related to the conformal anomaly on a curved spacetime
\begin{eqnarray}
\int_{S^4} \langle T_\mu^\mu\rangle.
\end{eqnarray}

In our model of the $SP(N)$ gauge theory,
the candidate of the conformal $R$ current
contains one free parameter $x=\gamma_Q$,
from which the corresponding $R$ charges
are determined by Eqs.(\ref{eq:fixedG1}) and (\ref{eq:fixedY1}) as
\begin{eqnarray}
R_Q = \frac{2}{3}(1+\frac{x}{2}), \quad
R_{Q'} = \frac{2}{3}(1+\frac{\gamma_{Q'}}{2}), \quad
R_{S} = \frac{2}{3}(1+\frac{-2x}{2})
\end{eqnarray}
with
\begin{equation}
\gamma_{Q'} = 1-\frac{3(N+1)-(N+1)(1-x)}{2(N+1)-\varepsilon},
\end{equation}
where $\varepsilon = 2(N+1)-n_F$. 

The claim is that among these one-parameter $R$ currents,
the conformal one maximizes the anomaly $a$,
which is obtained
as follows:
\begin{eqnarray}
a &=& 2N(2N+2)\left[3(R_Q-1)^3-(R_Q-1)\right] \cr
  & & +2(2N+2-\varepsilon)2N \left[3(R_{Q'}-1)^3-(R_{Q'}-1)\right] \cr
  & & + (N+1)(2N+1)\left[3(R_S-1)^3-(R_S-1)\right],
\label{a}
\end{eqnarray}
where we note that the $R$ charges appearing in $a$ are those of fermions
(i.e. $R_{\psi_{Q}} = R_{Q}-1$) because only fermions contribute to the anomaly.
By maximizing $a$ with respect to $x$, we can determine $x_{*}=\gamma_{Q}|_*$. The unique local maximum is achieved by setting 
\begin{eqnarray}
x_* &=& -\left(\varepsilon^2(2+3N)-4\varepsilon(1+N)(2+3N)
+(1+N)^2(8+13N)\right)^{-1} A; \cr
A &\equiv& 4-4\varepsilon+\varepsilon^2 + 22N-16\varepsilon N
 + 3 \varepsilon^2 N + 32N^2 - 12 \varepsilon N^2 + 14 N^3 
 + (\varepsilon - 2(1+N))B, \cr
B &\equiv& \sqrt{\varepsilon^2(1+2N)(1+6N)
-4\varepsilon (1+N)(1+2N)(1+6N)+ (2+9N+7N^2)^2}.
 \label{horrible}
\end{eqnarray}

To compare this rather complicated expression with the perturbative results,
we expand Eq.(\ref{horrible}) in terms of $\varepsilon$.
Remarkably, the first order approximation is given by 
\begin{equation}
x_1 = -\frac{N}{2+9N+7N^2}\varepsilon,
\end{equation}
which completely agrees with our Banks-Zaks-like calculation.
Furthermore we can systematically study higher order corrections.%
\footnote{For instance, the two loop contribution should be
\begin{eqnarray}
x_2 = -\frac{3N(1+N(7+11N))}{(1+N)^2(2+7N)^3}\varepsilon^2.
\nonumber
\end{eqnarray}
}
It is quite intriguing that the $a$-maximization determines
all-order loop effects only from the one-loop result Eq.(\ref{a}).

We can also study $a$ of the gauged version of extended IYIT model
in section 3,
which leads to conformal sequestering.
Since the gauging enforces yet another constraint on the anomaly free $R$
charge, we obtain a unique $R$ charge assignment without using the
$a$-maximization procedure. It is important to realize, following the
general argument of the monotonically decreasing $a$, that
$a_{\mathrm{gauged}}$ is less than $a_{\mathrm{ungauged}}$.
This is obvious when $x_*^{\mathrm{gauged}}$ is sufficiently close to
$x_*^{\mathrm{ungauged}}$, since $x_*^{\mathrm{ungauged}}$ yields the
local maximum of $a(x)$. For example, we can show by a direct
computation that $a$ of the gauged extended IYIT model presented in
section 3 is always less than that of the
ungaged version presented in section 2 for a fixed gauge group. 
This result is consistent with the fact that our conformal fixed point is a stable one. 
In particular, it is worthwhile to notice that this is even true for $N=1$ case, which cannot be treated in the one-loop approximation. 

Finally, it would be an interesting but challenging problem to obtain the
speed of the conformal sequestering from the interpolating $a$-function.
In Ref.\cite{Kutasov:2003ux},
the off-shell $a$-function is proposed as solving
$a$-maximization condition with a Lagrange multiplier
$\xi$ that enforces the constraint on the $R$ charge:
\begin{equation}
a(R(\xi),\xi) = \mathrm{Tr}\left(3 R^3 - R \right)+ \sum \xi (\mathrm{constraint}) \ ,
\end{equation}
where $R(\xi)$ is obtained by maximizing $a$ with respect to $R$
for fixed $\xi$, and the constraint is either ABJ anomaly free
condition or the requirement that the superpotential be marginal.
As was observed in \cite{Kutasov:2003ux},
the first derivative of $a(\xi)$ is related to the $\beta$ function
of the coupling constant.%
\footnote{Since the number of the Lagrange multipliers $\xi$
agrees with that of marginal deformations,
it is conjectured that $\xi$ can be regarded
as a coupling constant in a certain scheme.}
Furthermore, the second derivative (Hessian) of $a(\xi)$
at a fixed point $\xi_*$ is proportional to the slope of the $\beta$ function
\begin{equation}
\frac{\partial^2a(\xi)}{\partial \xi_i \partial \xi_j}\bigg|_*
 \propto \frac{\partial \beta_i (\xi)}{\partial \xi_j}\bigg|_*.
\end{equation}
 Consequently there is a chance to read the conformal sequestering matrix
 without performing the explicit loop calculation
 even for a strongly coupled theory.
 Unfortunately, we do not know the proportionality factor
 (related to the denominator of the NSVZ beta function evaluated
 at the fixed point) and the transformation matrix
 $\{{\partial \xi_i}/{\partial g_j}\}$
 non-perturbatively, so we cannot determine
 the conformal sequestering matrix.
 Since the conformal sequestering matrix
 is a physical renormalization invariant quantity
 while $a(\xi)$ is not,
 we need an off-shell scheme-independent $a$-function for our purposes.

\section{Another example of the hybrid scheme}
In section~{\ref{sec:Phen}}, we have considered the hybrid model of the anomaly and gauge mediated
SUSY breaking, which solves the tachyonic slepton problem in a pure anomaly mediation model.
In this appendix, we propose another example of the hybrid model which can be constructed with 
renormalizable interactions between the SUSY breaking sector and the messenger sector.
That is, in addition to the anomaly mediated SUSY breaking, we consider the gauge mediated SUSY
 breaking discussed in Ref.~\cite{Tobe},  where messenger sector consists of $N_m$ flavors of chiral
 superfields $\psi_i$ and  $\bar{\psi}_j$ $(i,j=1,\cdots N_m)$ and $N_m$ flavors of chiral superfields $\psi'_i$ and  $\bar{\psi}'_j$ $(i,j=1,\cdots N_m)$. Here, $\psi_i$, $\psi'_i$ and $\bar{\psi}_j$, 
 $\bar{\psi}'_j$ transform as ${\bf 5}$ and ${\bf 5^*}$ of $SU(5)$ GUT, respectively.
Then, with the superpotential, 
\begin{equation}
 hS_{ij}\psi_i\bar{\psi}_j + m_m \psi_i \bar{\psi}'_i + m_m \psi'_i \bar{\psi}_i,
 \label{eq:gmint}
\end{equation}
the SUSY-breaking effects are transmitted to the sfermions and Higgs bosons
by the gauge interactions.
Here, $h$ denotes the coupling constant,  $m_m$ the mass parameter, 
and we assume that $h = h_0$ of order one at $M_* \lsim M_G$. 
As discussed in section~\ref{sec:seq}, we impose the global $SU(2N+2)$ symmetry to the
SUSY breaking sector, and in order for the interaction in Eq.(\ref{eq:gmint}) to respect this 
symmetry, we assume that $N_m=2(N+1)$ and $\psi$, $\bar{\psi}$ and 
$\psi'$, $\bar{\psi}'$ transform as $\overline{\bf 2N+2}$
and $\bf{2N+2}$ 
representations, respectively, of the $SU(2N+2)$.%
\footnote{
Here, we assume $\vev{S}=0$, that is, $R$ symmetry is not broken.
In this case, gauginos do not obtain the SUSY-breaking masses via the gauge mediation
and the gaugino spectrum is the same as in the pure anomaly mediation.}

In this case, the scalar field $\phi$ obtains the mass squared via the gauge mediation,
and at the messenger scale, it is given by,
\begin{eqnarray}
m_{\phi}^2 &\simeq& 2 \sum_{a=1,2,3} C_a^{\phi}\bigg(\frac{\alpha_a}{4\pi}\bigg)^2 \frac{|h F_S|^2}{m_m^2},\\
&\simeq& 6 \sum_{a=1,2,3} C_a^{\phi}\bigg(\frac{\alpha_a}{4\pi}\bigg)^2
\bigg(\frac{m_{\rm phys}}{M_*}\bigg)^{\gamma_S}
\bigg(\frac{|h_0| M_G}{m_m}\bigg)^2
m_{3/2}^2.
\label{eq:GMmass2}
\end{eqnarray}
Here, we have used RG evolution of $hF_{S}$,
\begin{eqnarray}
h F_{S}\bigg|_{m_{m}} =h F_{S}\bigg|_{m_{\rm phys}} = 
\bigg(\frac{m_{\rm phys}}{M_*}\bigg)^{\frac{\gamma_S}{2}} h_{0} F_{S} \bigg|_{m_{\rm phys}} 
\simeq \bigg(\frac{m_{\rm phys}}{M_*}\bigg)^{\frac{\gamma_S}{2}} h_{0} m_{3/2}M_{G},
\end{eqnarray}
where subscripts $(m_{m}, m_{\rm phys})$ denote the RG scale.
Furthermore, in the course of RG running, the scalar mass squared is suppressed by
a factor of $\eta$ which denotes the sequestering effects between the scales 
$m_m$ and $m_{\rm phys}$,
\begin{eqnarray}
\eta = 
\left\{
\begin{array}{ccc}
(m_{\rm phys}/m_m)^{\beta'_*}  &  {\rm for} &m_{m} > m_{\rm phys},\\ 
 1 &   {\rm for} &m_{m} \leq m_{\rm phys}.
\end{array}
\right.
\end{eqnarray}

For example, for the $SP(3)\times SP(1)^2$ model, 
the positive contribution to the slepton masses squared in Eq.(\ref{eq:GMmass2}) is given by
\begin{eqnarray}
m_{\tilde{e}\,\, \rm G.M.}^2 \simeq  \frac{3}{5}\,\, 6\, 
\bigg(\frac{m_{\rm phys}}{m_m}\bigg)^{2}
\bigg(\frac{M_G}{M_*}\bigg)^{2}
\bigg(\frac{\alpha_1}{4\pi}\bigg)^2m_{3/2}^2 
\eta,
\end{eqnarray}
where we are using $\gamma_S = 2$ and are assuming $h_0=1$.%
\footnote{Here, we have neglected RG effects from the MSSM couplings.}
Thus, when $m_{m}$ satisfies 
\begin{eqnarray}
m_{m} \lsim
\left\{
\begin{array}{ccc}
m_{\rm phys} (M_G/M_*)^{2/(2+\beta_*')} &  {\rm for} &m_{m} > m_{\rm phys},\\ 
 m_{\rm phys} (M_G/M_*)  &   {\rm for} &m_{m} \leq m_{\rm phys},
 \end{array}
\right.
\end{eqnarray}
this contribution overcomes the negative contribution from the anomaly mediated mass squared
in Eq.(\ref{eq:AMse}).%
\footnote{In this model, the gauge coupling constants in the SSM remain perturbative up to 
Grand Unification scale $M_{\rm GUT}\simeq 2\times 10^{16}$~GeV for $m_m\gsim 10^{13}$~GeV.
}
Therefore, we find that this hybrid model also provides a solution to the tachyonic slepton problem 
with an appropriate choice of the mass scale of messengers.%
\footnote{Since the messenger particles are very heavy $m_m \gsim 10^{13}$~GeV, they are not
 produced thermally for the reheating temperature of the universe around 
 $T_R \simeq 10^{10}$~GeV, which is very advantageous for the thermal 
 leptogenesis~\cite{thermalLG}.}

\end{document}